\documentclass[12p]{iopart} 
\usepackage{hyperref}
\usepackage{graphics} 
\usepackage{graphicx} 
\usepackage{subfigure}
\usepackage{url}
\usepackage{tipa}
\usepackage{iopams}
\usepackage{braket}
\usepackage{todonotes}
\usepackage{cancel}
\usepackage{cite}

\begin{document} 
\article{}{Reversible decay of ring dark solitons}
\author{L A Toikka and K-A Suominen}
\address{Turku Centre for Quantum Physics and Laboratory of Quantum Optics, Department of Physics and Astronomy, University of Turku, 20014 Turku, Finland}
\ead{\mailto{laantoi@utu.fi}, \mailto{kalle-antti.suominen@utu.fi}}

\begin{abstract}
We show how boundary effects can cause a Bose-Einstein condensate to periodically oscillate between a (circular) array of quantised vortex-antivortex pairs and a (ring) dark soliton. If the boundary is restrictive enough, the ring dark soliton becomes long-lived.
\end{abstract}

\maketitle

\section{\label{sec:Introduction}Introduction}
The method of images~\cite{Donnelly91} has been successfully applied~\cite{PhysRevA.60.R1779,PhysRevA.77.032107,PhysRevA.64.033607,PhysRevA.74.043611} to the dynamics of quantised vortices in superfluids and Bose-Einstein condensates~\cite{P&S}, including on toroidal geometry~\cite{PhysRevA.66.033602,PhysRevA.64.063602,PhysRevA.79.043620}. In the absence of external forces, a vortex moves at the external superfluid velocity, which is given by an appropriate configuration of image vortices. In particular, the image system of a vortex line of charge $\kappa$ in the $z$-direction next to a rigid plane $y = 0$ is an equal and opposite vortex at the mirror image~\cite{Saffman97}. The induced velocity field is $\kappa / (4\pi h)$ in the $y$-direction, where $h$ is the distance from the wall. The vortex moves parallel to the surface, but the limit $h \to 0$ is not defined. 

Near to the wall the velocity of the vortex exceeds the Landau critical velocity~\cite{Landau41} $v_\mathrm{L}$, above which the energy can be lowered by generating dissipative excitations on top of the superfluid component, and the vortex decays through the creation of phonons. In the homogeneous case, $v_\mathrm{L}$ equals to the speed of sound $v_\mathrm{s}$, but in the presence of defects that strongly alter the density of the condensate, $v_\mathrm{L}$ can be reduced by more than a factor of two~\cite{PhysRevLett.69.1644}.

If, however, instead of a single boundary we place another one at the same distance onto the other side of the vortex, by symmetry the net external velocity due to the images becomes zero, and the vortex remains stationary. This observation poses the interesting (open) question that what happens when the condensate becomes narrow, especially when it supports a one-dimensional array of vortex-antivortex pairs? 

It is well known that such an array is produced by the decay of a dark soliton~\cite{Kivshar1998,Emergent,Greekreview} in two dimensions through the snake instability~\cite{PhysRevA.88.063610,PhysRevA.87.043601,PhysRevA.60.R2665,PhysRevA.62.053606,PhysRevA.64.063602,ZacharyDutton07272001, PhysRevLett.86.2926}, but the ensuing vortex dynamics are not so well studied. In systems with cylindrical symmetry, also ring dark solitons (RDSs)~\cite{PhysRevE.50.R40,PhysRevLett.90.120403} have been observed to decay into a circular array of vortex-antivortex pairs~\cite{PhysRevA.87.043601}, often called a necklace. To date, RDSs have not been observed with cold atoms, but proposals for their experimental creation have been very recently presented~\cite{0953-4075-47-2-021002,Toikka2013}. Furthermore, exact soliton-like solutions of the cylindrically symmetric Gross-Pitaevskii equation corresponding to a dark ring, which also exhibit the snake instability, are known~\cite{ths2012}. In~\cite{PhysRevA.87.043601,0953-4075-47-2-021002}, it was shown that revival of the original RDS(s) is possible, and in this work we focus on the details of the vortex recombination. In~\cite{Shomroni09}, it was experimentally observed that a short planar dark soliton can be periodically revived from a single vortex-antivortex pair it decays into.

In the work presented here, in particular, we theoretically show that the snake instability is not always irreversible, and predict that the original (ring) dark soliton is revived when the width $d$ of the condensate in the perpendicular direction to the original soliton is $d \lesssim 16 \xi$, where $\xi$ is the healing length. In the case of a RDS, $d$ is the radial width of the annular trap. We show that the revival occurs periodically, i.e. the condensate oscillates between a RDS state and a vortex-antivortex necklace state, when the two states are near in energy and number of atoms. If $d$ is decreased further ($d \lesssim 3 \xi$), we show that the RDS becomes long-lived, with lifetimes$~\mathcal{O}(1\,\mathrm{sec})$.

\section{Theoretical background}
In this work, we consider a scalar order parameter $\psi$, representing the macroscopic wavefunction of a Bose-Einstein condensate trapped in a potential given by $V_{\mathrm{trap}}$, which is a solution to the Gross-Pitaevskii equation:
\begin{equation}
\label{eqn:nls0}
\rmi \psi_t = -\nabla^2 \psi + V_{\mathrm{trap}}\psi + C_{\mathrm{2D}}|\psi^2|\psi.
\end{equation}
Here we have assumed a two-dimensional condensate whereby the $z$-direction is tightly trapped to the corresponding harmonic oscillator ground state ($\omega_z \gg \omega_x = \omega_y \equiv \omega_r$) and has been projected onto the $xy$-plane. Then $C_{\mathrm{2D}} = 4 \sqrt{\pi} Na/a_{\mathrm{osc}}^{(z)}$, where $N$, $a$, and $a_{\mathrm{osc}}^{(z)}$ are the number of atoms in the cloud, the $s$-wave scattering length of the atoms, and the characteristic trap length in the $z$-direction respectively. We have obtained dimensionless quantities by measuring time, length and energy in terms of $\omega_r^{-1}$, $a_{\mathrm{osc}}^{(r)} \equiv a_{\mathrm{osc}} = \sqrt{\hbar/(2m\omega_r)}$ and $\hbar \omega_r$ respectively, where $\omega_r$ is the angular frequency of the trap in the $r$-direction. This basis is equivalent to setting $\omega_r = \hbar = 2m = 1$. All the units in this work are expressed in this dimensionless basis. Also, we denote $\rho \equiv |\psi|^2$.

\section{Periodic ring dark soliton/vortex-antivortex necklace}

\subsection{\label{sec:snake}Snake instability}
In two dimensions, the snake instability means the decay of a (ring) dark soliton into a (circular) array of vortex-antivortex pairs. In three-dimensional systems, the snake instability results in vortex rings. In~\cite{PhysRevA.87.043601}, it was shown that the snake instability follows after a symmetry breaking; subsequently an unbalanced quantum pressure on opposite sides of the (ring) dark soliton notch starts generating the 'snaking' behaviour. Furthermore, the number of vortex-antivortex pairs $N_{\mathrm{v}}$ after the snake instability of a RDS of radius $R$ was shown to be approximately 
\begin{equation}
\label{eqn:N_v}
N_{\mathrm{v}} \sim \frac{2\pi R}{8\xi}.
\end{equation}

\subsection{\label{sec:recomb}Vortex recombination as a boundary effect}
\begin{figure}
\centering
\includegraphics[width=\textwidth] {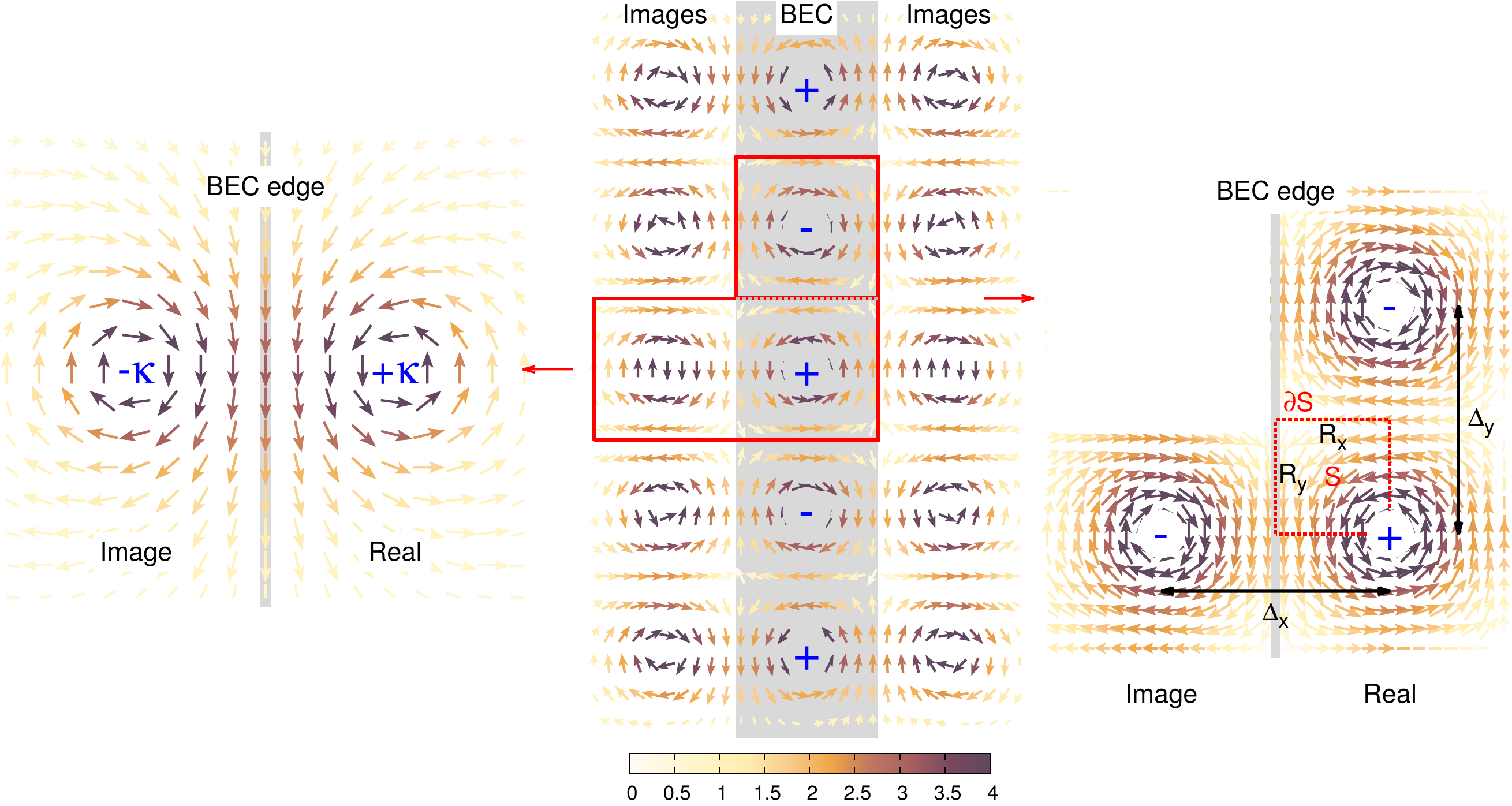}
\caption{\label{fig:dipole_streamlines} Superfluid flow lines for a narrow condensate with a vortex-antivortex chain of alternating charges $\pm \kappa$ (blue labels) as given by the method of images. The BEC edges impose boundary conditions that restrict the flow. The region inside the red box is shown on the left. On the right, a region with three vortices.}
\end{figure}

In the absence of external forces such as dissipation, a quantised vortex of circulation $\kappa_i$ at $\bi{r}_i$ moves with the condensate flow at the external superfluid velocity $\bi{v}_{\mathrm{F}}(\bi{r}_i,t)$~\cite{Saffman97}. The total superfluid velocity field $\bi{v}(\bi{r},t)$ is then
\begin{equation}
\label{eqn:vortexflow}
\bi{v}(\bi{r},t) = \frac{\kappa_i}{2\pi}\frac{\hat{\bi{z}}\times(\bi{r}-\bi{r}_i)}{|\bi{r}-\bi{r}_i|^2} + \bi{v}_{\mathrm{F}}(\bi{r},t),
\end{equation}
and the vortex motion is given by 
\begin{equation}
\label{eqn:vortexflow1}
\frac{\mathrm{d}\bi{r}_i}{\mathrm{d}t} = \bi{v}_{\mathrm{F}}(\bi{r}_i,t).
\end{equation}

Let a single vortex of charge $\kappa$ be at the origin, $\bi{r}_i = (0,0)$, and let us consider a rigid wall at $x = -\Delta_x/2$. To enforce the kinematic boundary condition $\bi{v} \cdot \hat{\bi{n}} = -v_x = 0$, where $\hat{\bi{n}}$ is the outward unit normal vector, we add an image vortex of circulation $-\kappa$ at $(x,y) = (-\Delta_x,0)$. Using~\Eref{eqn:vortexflow}, the velocity field is then (see~\Fref{fig:dipole_streamlines} left panel)
\begin{equation}
\label{eqn:v_dip}
\bi{v}(x,y) = \frac{\kappa}{2\pi} \left( \begin{array}{cc} \frac{y}{(\Delta_x +x)^2+y^2}-\frac{y}{x^2+y^2} \\ \frac{x}{x^2+y^2} - \frac{\Delta_x +x}{(\Delta_x +x)^2+y^2} \end{array} \right).
\end{equation}
Similarly, to consider a longer array of real vortices separated by $\Delta_y$ in a narrow condensate, we may add more images (see~\Fref{fig:dipole_streamlines} middle panel), and it is a good approximation to neglect any further images required by the images themselves. It is clear by symmetry that for every real vortex in the (infinite) array, the vector sum of the external fields generated by all the images and the other real vortices is zero, and the vortex-antivortex array is stationary. The periodic boundary condition of annular geometry results in an infinite one-dimensional array (a necklace) of vortex-antivortex pairs, and by cylindrical symmetry it suffices to consider only a single one. On the other hand, the inner and outer rims (with radii $R_1$ and $R_2$ respectively) generate an infinite set of images for each real vortex~\cite{Saffman97}, but if $R_1/R_2 \sim 1$ in the limit as $R_1,R_2 \to \infty$, the situation is equivalent to the infinite straight array. Besides, as we will show below, a narrow torus is needed for the vortex recombination, so we focus on this limit.

\begin{figure}
\centering
\includegraphics[width=0.45\textwidth] {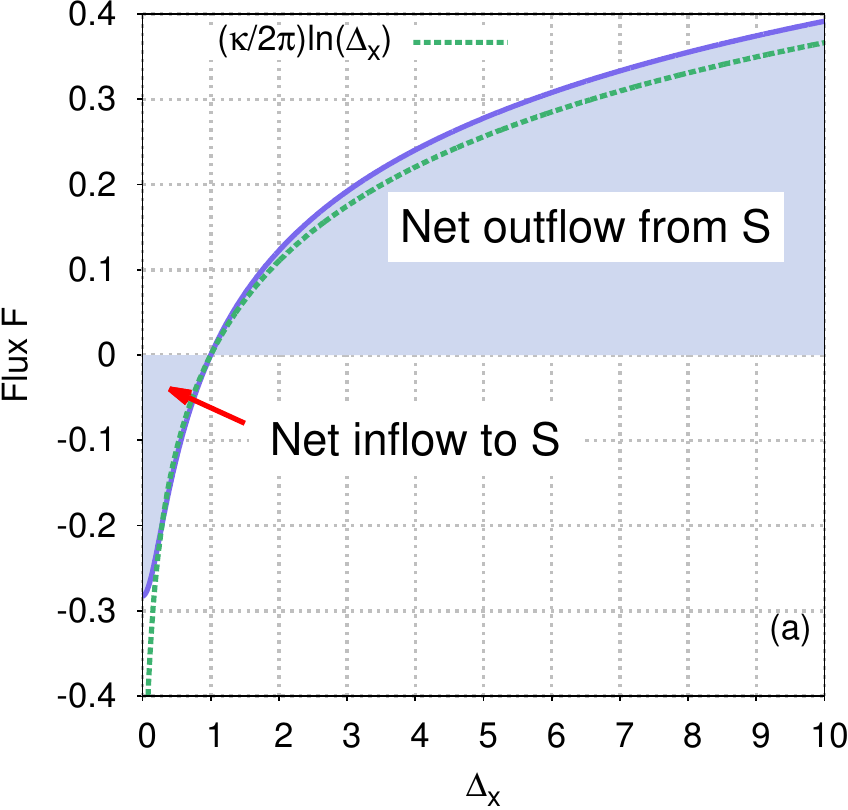}
\includegraphics[width=0.45\textwidth] {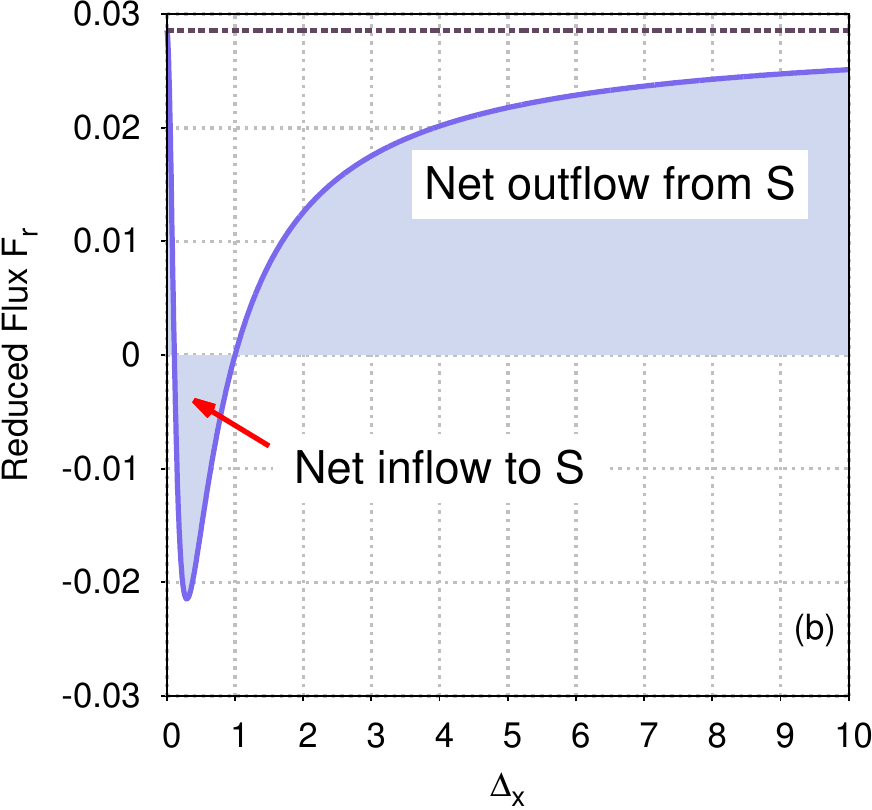}
\caption{\label{fig:flux} (a) The flux $F$ (see~\Eref{eqn:surfint}) normalised to $n_0 = 1$ through the surface $\partial S$ (solid line). The flux diverges logarithmically for large $\Delta_x$ (dashed line). (b) The reduced flux $F_{\mathrm{r}}$ (see~\Eref{eqn:Fr}), obtained by dropping the logarithmic terms in $F$. The dashed horizontal line shows the asymptotic limit of $F_{\mathrm{r}}$ as $\Delta_x \to \infty$ (see~\Eref{eqn:Fr_lim}). We use $\Delta_y = 1$, and $\xi = 0.1$, which roughly correspond to the evolution in~\Fref{fig:revival}.}
\end{figure}
The superfluid velocity field~\eref{eqn:v_dip} diverges at the vortex cores, so we invoke a variational approximation for the vortex density profile~\cite{P&S}, 
\begin{equation}
\label{eqn:nprofile}
n(r) = n_0 \frac{r^2}{2\xi^2 + r^2},
\end{equation}
where $n_0$ is the constant bulk density and $\xi = 1/\sqrt{8\pi n_0 a}$ is the healing length\footnote{Using~\Eref{eqn:nprofile}, the distance $r_0$ from the vortex core beyond which the flow speeds are less than the speed of sound $v_\mathrm{s}(\textbf{r}) = \sqrt{2n(\bi{r})C_{\mathrm{2D}}}$ is given by $r_0 = \left(\frac{2\kappa^2 a}{\pi C_{\mathrm{2D}}}\right)^{\frac{1}{4}}\xi \ll \xi$.}. Thus, integrating the continuity equation $\dot{\rho}(\bi{r},t) = -\nabla \cdot [\rho(\bi{r},t) \bi{v}(\bi{r})]$ around the closed path $\partial S$ (see~\Fref{fig:dipole_streamlines} right panel) such that $R_x = \Delta_x/2$ and $R_y = \Delta_y/2$, we get
\begin{eqnarray}
\label{eqn:surfint}
F & \equiv -\int_S \frac{\partial \rho(\bi{r},t)}{\partial t}\, \mathrm{d}\it{A} = \int_S \nabla \cdot [n(r)\bi{v}(\bi{r})]\, \mathrm{d}A \\
&= \oint_{\partial S} n(r)\bi{v}(\bi{r}) \cdot \mathrm{d}\hat{\bi{n}} \\
\nonumber
&=  \frac{\kappa}{2\pi}\frac{n_0 \Delta_y^2}{2\Delta_y^2 + 4\xi^2}\left[ \frac{\sqrt{8}\xi}{\Delta_y} \arctan{\left( \frac{\Delta_y}{\sqrt{8}\xi} \right)} \right. \\
\label{eqn:surfint1}
& \left. \qquad + \ln{\left(\frac{2\xi^2}{\Delta_y^2 + 8\xi^2} \right)} \right]  \\
\nonumber
&\,\,\,\,\,\, - (\Delta_y \leftrightarrow \Delta_x).
\end{eqnarray}

When $\Delta_y = \Delta_x$, there is no net density flow into or out of $S$ (see~\Fref{fig:flux}). In fact, by symmetry even the stricter local condition $\nabla \cdot [n(r)\bi{v}(\bi{r})] = 0$ at every $\bi{r} \in S$ is satisfied. We also note that since $\lim_{r\to\infty}{n(r)} = n_0$, $F$ diverges logarithmically at large $\Delta_x$ or $\Delta_y$ (unless $\Delta_x,\Delta_y \gg 1$ and $\Delta_x \sim \Delta_y$). The divergence is similar to the one encountered when using a logarithmic potential to calculate the electrostatic field due to an infinite line charge, as such a potential does not allow for a zero boundary condition at infinity. In a physical BEC, the density goes to zero at the cloud edges. We obtain a reduced flux $F_{\mathrm{r}}$ by dropping the logarithmic terms in $F$;
\begin{equation}
\label{eqn:Fr}
\eqalign{
F_{\mathrm{r}} = \frac{\kappa n_0}{2\pi}\frac{ \sqrt{2}\xi \Delta_y}{\Delta_y^2 + 2\xi^2} \arctan{\left( \frac{\Delta_y}{\sqrt{8}\xi} \right)} 
 \nonumber \cr
 \qquad - (\Delta_y \leftrightarrow \Delta_x).}
\end{equation}

\begin{figure}
\centering
\includegraphics[width=\textwidth] {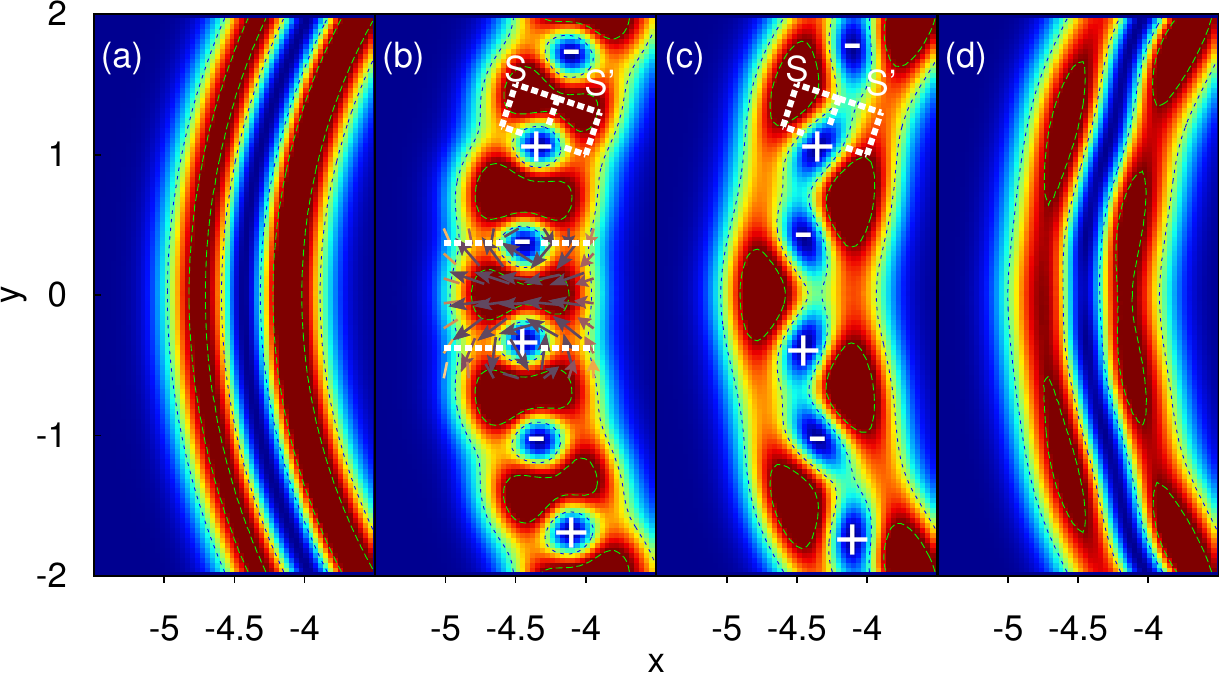}
\caption{\label{fig:revival} \emph{Revival of the RDS.} Density plots of the snake instability and subsequent revival of the RDS as given by the 1RDS case in~\cite{0953-4075-47-2-021002}. Here $\xi = 0.1$, measured as in~\Sref{sec:annwidth} (a) Initially the condensate contains the RDS. (b) The snake instability has transformed the RDS into a vortex-antivortex necklace. The vortex charges are shown in white. The overlaid vector field shows the superfluid velocity (see~\Eref{eqn:v_dip}), and the horizontal white dashed lines show where the flow is partially blocked as discussed in~\Sref{sec:recomb}. Examples of the areas $S$ and $S'$ are also shown. (c) The partially blocked flow results in the condensate accumulating in $S$ away from $S'$, causing a depletion of the density between the vortices. (d) The RDS is revived. The colouring is the same in each figure, and the times are (a) $t = 0$, (b) $t = 1.6$, (c) $t = 1.8$, (d) $t = 2.1$. }
\end{figure}

The main contribution to a non-zero $F$ is having $\Delta_y \neq \Delta_x$. For $\Delta_x < \Delta_y$, $F < 0$, which means that net density is deposited inside $S$ (see~\Fref{fig:flux}), i.e. the flow lines that pass near to the wall cannot support a sufficient return flux from above the vortex. The result is a depletion of the density along the line connecting the vortex to the antivortex below it, and a dark soliton is formed between them, as confirmed by direct numerical integration of the GPE~\eref{eqn:nls0} (see~\Fref{fig:revival}, and also~\Sref{sec:annwidth}). Similarly, if the area $S$ is placed to the right of the vortices (let us call it $S'$), then $F > 0$, and there will be a net outflow of density away from $S'$ (see~\Fref{fig:revival}(b),(c)).

In the special case that the condensate in~\Fref{fig:revival}(d) would be identical to that in~\Fref{fig:revival}(a), the decay and revival dynamics would be periodic. However, the density is slightly altered, and as a result, the vortices will not be uniformly distributed along the torus after the second decay into a vortex-antivortex necklace. This deviation will cause the subsequent vortex-antivortex necklaces to differ in energy and number of atoms from the original RDS, and further revivals become less likely.

\subsection{\label{sec:numres}Results for various values of $\Delta_x/\Delta_y$.}

\subsubsection{Dipole annihilation.}
\label{sec:dipannih}
\begin{figure}
\centering
\includegraphics[width=0.45\textwidth] {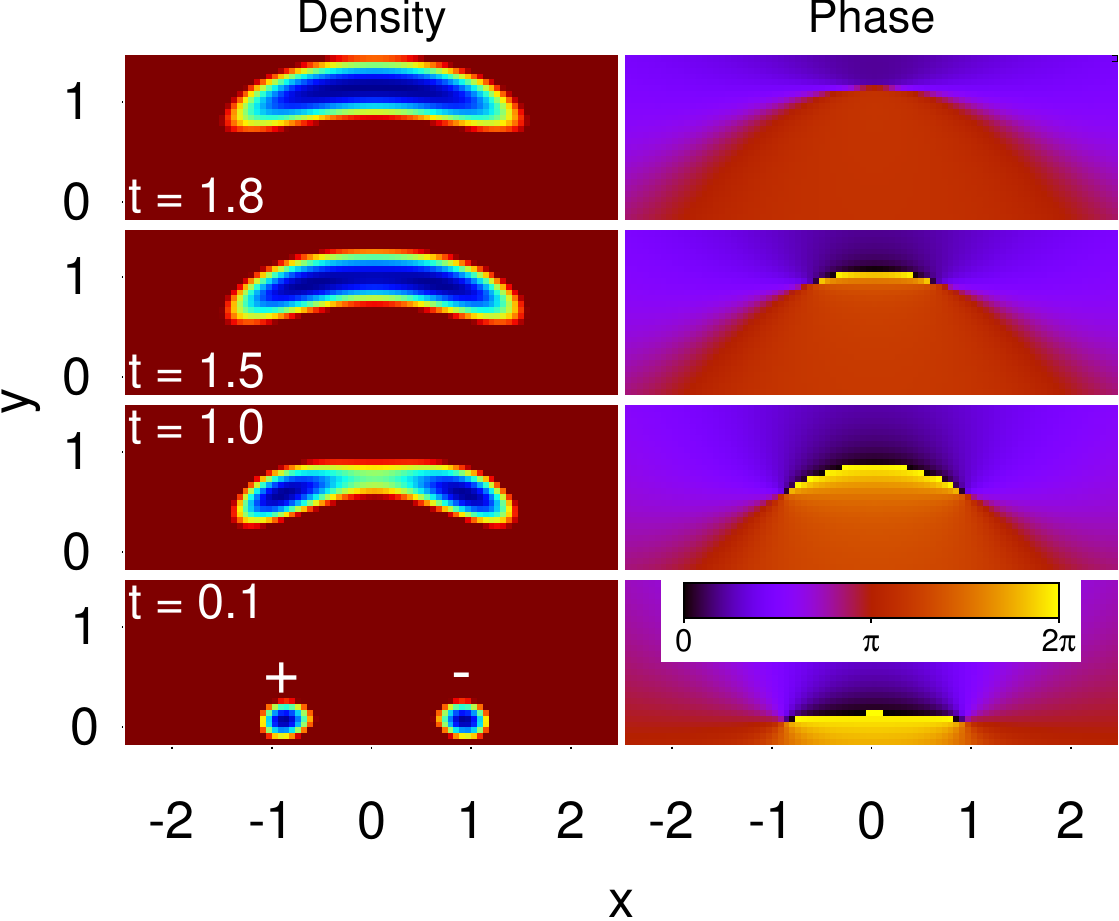}
\caption{\label{fig:dipannih} If the initial separation is small enough, the vortices of a lone dipole far from boundaries close in and annihilate each other, producing a short grey soliton. See the text for discussion. Here $\xi \approx 0.8$, obtained by finding a best fit to~\Eref{eqn:nprofile} along a slice $x = \pm 1$ in the density at $t = 0$, making the dipole separation $\sim 2.5\xi$. In the simulation we use $C_{\mathrm{2D}} = 50$.}
\end{figure}

\Eref{eqn:Fr} allows us to probe different limits for $\Delta_x$ and $\Delta_y$. For example, having $\Delta_x \gg 1$ and $\Delta_y \sim 1$ corresponds to a lone vortex-antivortex pair (also known as a vortex dipole~\cite{PhysRevA.76.043602,PhysRevA.68.063609}) away from boundaries. As shown in~\Fref{fig:flux}(b), the region $S$ ($S'$) will lose (gain) density, and so our result~\eref{eqn:Fr} predicts that the vortices in such a lone dipole will move diagonally towards each other in the direction of the superfluid flow between them. By~\Eref{eqn:vortexflow1}, the speed of the dipole (in the $x$-direction) is given by the flow field of one vortex at the location of the other, which is proportional to $\Delta_y^{-1}$. Therefore, in general, such a lone dipole will accelerate and become closer until the vortices of the dipole overlap and annihilate each other. This result is confirmed by a numerical integration of the GPE~\eref{eqn:nls0} (see~\Fref{fig:dipannih}). We have assumed that the initial $\Delta_y$ is small enough; the limit
\begin{eqnarray}
\label{eqn:Fr_lim}
\lim_{\Delta_x \to \infty} F_{\mathrm{r}} &= \frac{\kappa n_0}{2\pi}\frac{ \sqrt{2}\xi \Delta_y}{\Delta_y^2 + 2\xi^2} \arctan{\left( \frac{\Delta_y}{\sqrt{8}\xi} \right)},
\end{eqnarray}
which amounts to dropping the $\Delta_x$ term in $F_{\mathrm{r}}$, shows that if in addition $\Delta_y \to \infty$, we also drop the $\Delta_y$ term and the effect of the other vortex vanishes, as might be intuitively expected. Also, the limit $\xi \to 0$ has the same effect in~\Eref{eqn:Fr_lim} (that the influence of the other vortex becomes negligible, see subsection~\ref{sec:annwidth}).

\subsubsection{\label{sec:annwidth}Width of the annulus.}
\begin{figure}
\centering
\includegraphics[width=\textwidth] {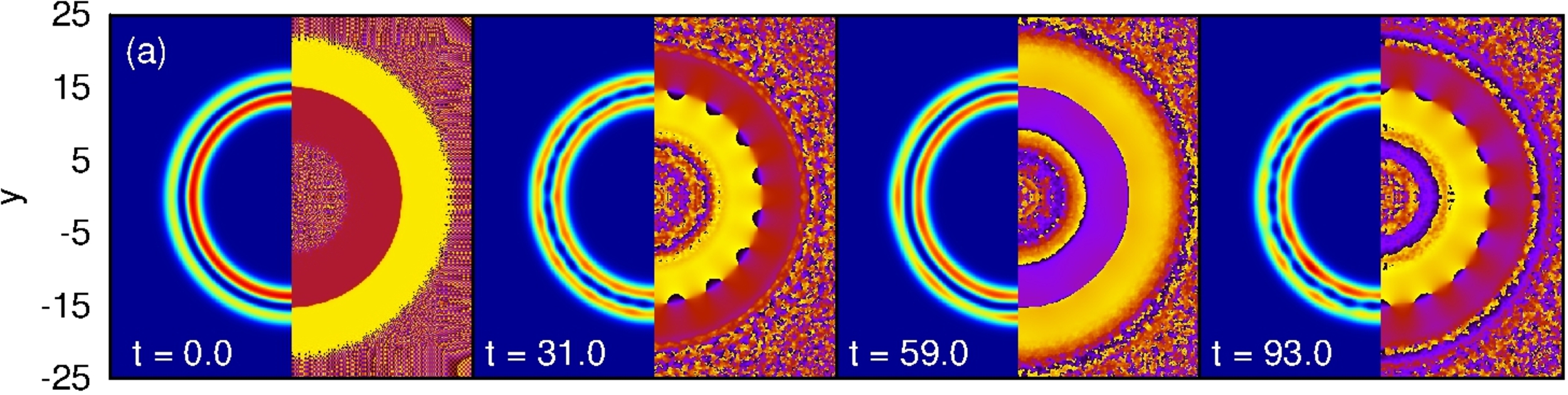}\\
\includegraphics[width=\textwidth] {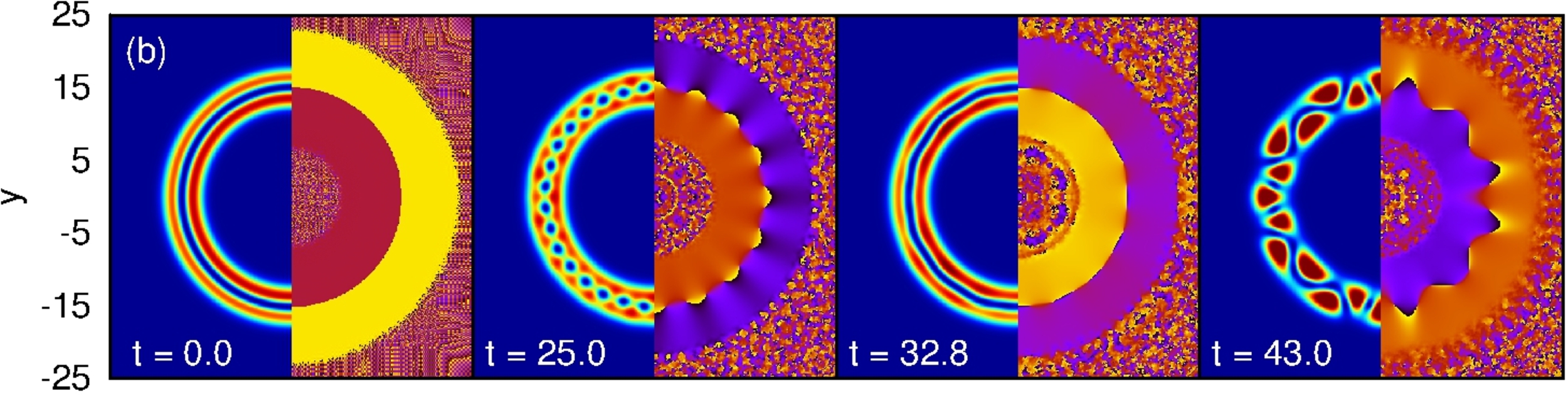}\\
\includegraphics[width=\textwidth] {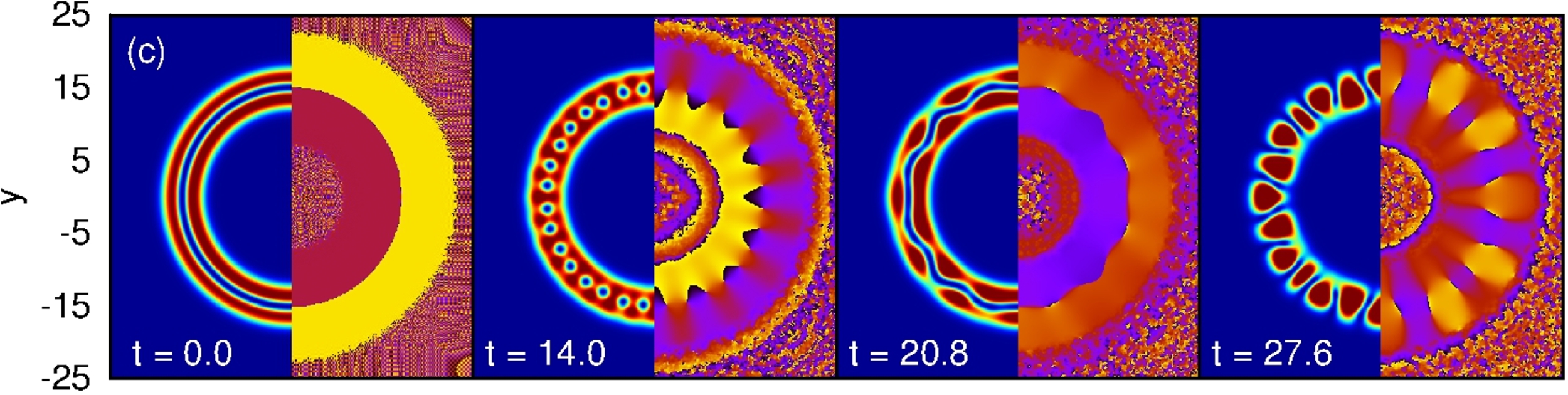}\\
\includegraphics[width=\textwidth] {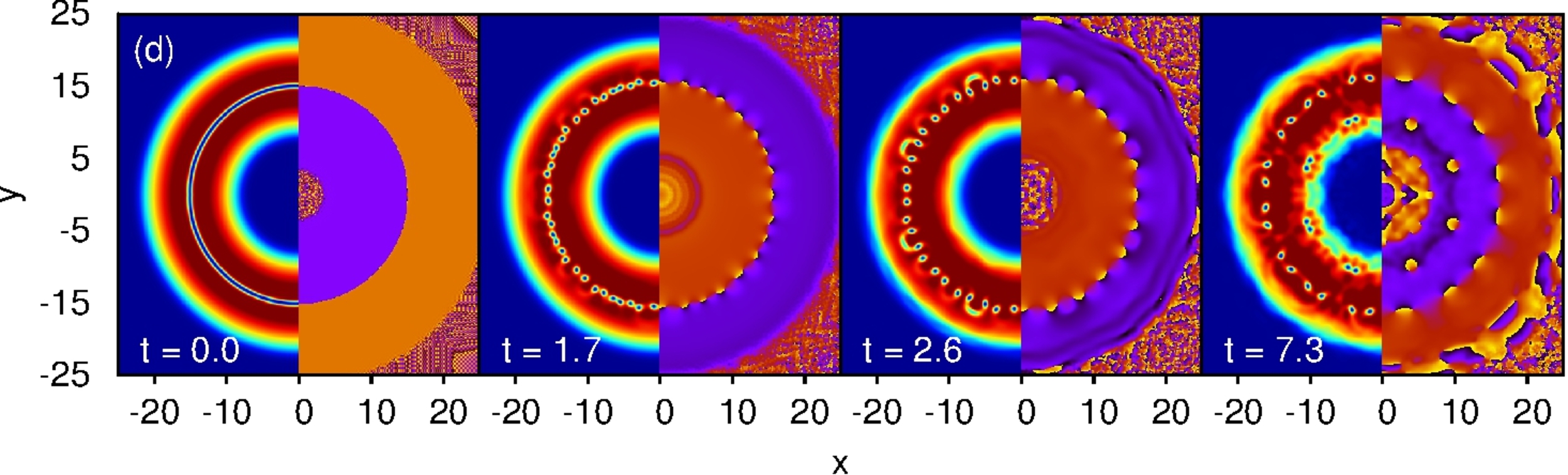}\\
\caption{\label{fig:annwidth} The symmetrical density (left) and phase (right) distributions at various times in the time-evolution as given by the GPE~\eref{eqn:nls0}. Revival (third column from the left) of the initial RDS (first column) after the snake instability (second column) depends on the width of the annulus (see~\Tref{tab:1} and the text for discussion). In (a), there also occurs a second almost full revival of the already revived RDS at $t = 107.0$. In (c), not all the vortex-antivortex pairs completely disappear at $t = 20.8$, and the revival is partial. However, the smaller $\Delta_x/\Delta_y$ is, the more complete (and also slower) the revival becomes. In (d) at $t = 7.3$, a few closely enough separated vortex dipoles have travelled inwards, in accordance with the mechanism described in~\Sref{sec:dipannih}. The dipoles split at the inner rim eventually swopping pairs and forming new dipoles that accelerate outwards, which annihilate before reaching the outer rim. The density colouring is not to scale amongst the subfigures, and the phase colouring is the same as in~\Fref{fig:dipannih}. (a) $C_{\mathrm{2D}} = 200$. (b) $C_{\mathrm{2D}} = 350$. (c) $C_{\mathrm{2D}} = 500$. (d) $C_{\mathrm{2D}} = 12500$.}
\end{figure}
\begin{table}
\caption{\label{tab:1}The chemical potential ($\mu$), healing length ($\xi$), and the vortex-antivortex necklace aspect ratio ($\Delta_x/\Delta_y$) for various nonlinearities ($C_{\mathrm{2D}}$). The potential is given by $V_{\mathrm{trap}}(r) = \frac{1}{4}(r-15)^2$. 'Revival' indicates the reversibility of the snake instability, i.e. all point phase singularities disappear.}
\begin{indented}
\item[]\begin{tabular}{@{}c|cccc}
\br
$C_{\mathrm{2D}}$&$\mu$&$\xi$&$\Delta_x/\Delta_y$&Revival\\
\mr
100&1.71&1.50&0.44&yes\\
200&1.92&1.10&0.63&yes\\
350&2.21&0.85&0.87&yes\\
400&2.30&0.80&0.95&yes\\
500&2.49&0.75&1.05&partial$^{\rm a}$\\
1000&3.33&0.60&1.52&no\\
%1300&3.79&0.55&1.76&no\\
%1600&4.22&0.50&2.06&no\\
%1900&4.62&0.47&2.28&no\\
%2100&4.88&0.46&2.40&no\\
2500&5.38&0.44&2.64&no\\
%5000&8.07&0.35&4.06&no\\
%7500&10.33&0.31&5.58&no\\
%10000&12.36&0.28&6.28&no\\
12500&14.22&0.26&7.26&no\\
50000&34.90&0.17&17.38&no\\
\br
\end{tabular}
\item[] $^{\rm a}$ See~\Fref{fig:annwidth}(c).
\end{indented}
\end{table}

From~\Eref{eqn:N_v}, we can obtain an estimate of $\Delta_y \approx 8\xi$ for the vortex-antivortex separation in the necklace after the snake instability of a RDS, agreeing with~\Fref{fig:revival} and $\Delta_y = 1$, $\xi = 0.1$ as used in~\Fref{fig:flux}. Therefore, our results predict that the revival dynamics of the RDS occurs when $d \lesssim 16 \xi$, where $d$ is the radial width of the annulus. We assume that the RDS remains stationary irrespective of $d$ and equally distant from both edges.

The healing length $\xi$, and hence the relative magnitudes of $\Delta_x$ and $\Delta_y$ can be controlled by changing the nonlinearity $C_{\mathrm{2D}}$ whilst keeping the potential $V_{\mathrm{trap}}$ fixed. We integrate numerically the GPE~\eref{eqn:nls0} using a range of values for $C_{\mathrm{2D}}$ (see~\Tref{tab:1} and~\Fref{fig:annwidth}). As an initial state for every run (i.e. value of $\xi$) we print a RDS at the radius $R_{\mathrm{S}}$, which produces a vortex-antivortex necklace through the snake instability, and the trapping potential is given by $V_{\mathrm{trap}} = \frac{1}{4}(r-R_{\mathrm{S}})^2$. We choose $R_{\mathrm{S}} = 15$, and monitor the subsequent vortex dynamics. The healing length $\xi$ is estimated by finding a best fit to $\frac{\mu-\frac{1}{4}(r-R_{\mathrm{S}})^2}{C_{\mathrm{2D}}} \tanh^2{\left(\frac{r-R_{\mathrm{S}}}{\sqrt{2} \xi}\right)}$ of the initial RDS state, where $\mu$ is the chemical potential, and $\Delta_x$ is estimated by the same Thomas-Fermi approximation to obtain $\Delta_x \approx 4\sqrt{\mu}$, giving
\begin{equation}
\label{eqn:aspratio}
\frac{\Delta_x}{\Delta_y} = \frac{\sqrt{\mu}}{2\xi}.
\end{equation}

As we can see in~\Tref{tab:1}, the initial RDS is revived in the numerically obtained time-evolution of the GPE~\eref{eqn:nls0} when $\Delta_x \leq \Delta_y$, which agrees with the theoretical result of~\Sref{sec:recomb} (see~\Fref{fig:flux}). Therefore, using~\Eref{eqn:aspratio}, the vortex-antivortex necklace recombines back to the RDS if $\sqrt{\mu} \lesssim 2\xi$. This result is a special case of the more general condition $d \lesssim 16 \xi$ (see above) for $V_{\mathrm{trap}} = \frac{1}{4}(r-R_{\mathrm{S}})^2$.

We note that for $C_{\mathrm{2D}} = 100$, the initial RDS is long-lived. In our simulation, the onset of the snake instability happens at $t \approx 223$, and the (full) revival at $t = 263$. The prolonged stability as $\xi$ becomes larger is in line with the discussion in~\cite{PhysRevA.87.043601}. Using the experimental parameters of the toroidal BEC of~\cite{PhysRevLett.111.205301} and $^{87}\mathrm{Rb}$, we obtain $a_{\mathrm{osc}} = 0.8\, \mu\mathrm{ m}$, $C_{\mathrm{2D}} \sim 200-2000$, and one unit of time is $\sim 1.5$ ms. Therefore, the life-time of the RDS is $\approx 335\,\mathrm{ms}$. Decreasing $C_{\mathrm{2D}} = 100$ further will increase the life-time, but we have not performed exhaustive simulations in this weakly-interacting limit.

\subsection{What if all the vortices are of equal sign?} 
\begin{figure}
\centering
\includegraphics[width=0.45\textwidth] {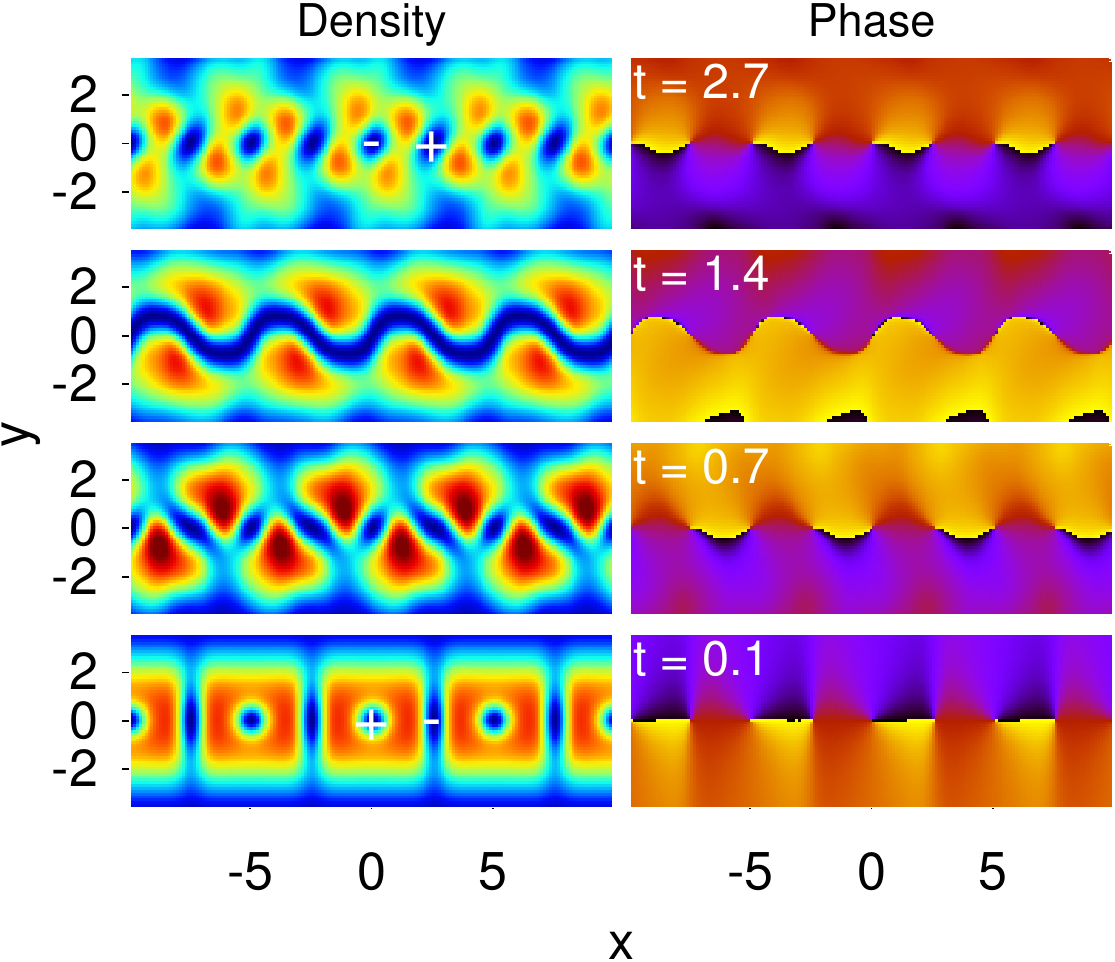}
\caption{\label{fig:samesigns} An infinite array (periodic boundary conditions apply in the $x$-direction) of identical vortices develops into an alternating chain of vortices and antivortices, which then combines to produce a dark soliton. After the soliton decays, the vortices have swopped charges. Phase colouring is the same as in~\Fref{fig:dipannih}.}
\end{figure}
If all the real vortices have a positive charge of $\kappa = 1$ in the infinite array (see~\Fref{fig:dipole_streamlines} middle panel), the effect on motion of a single real vortex by all the images and the other real vortices still cancels by symmetry, and so by the image argument the vortices should remain stationary. However, an initial state consisting of a periodic array of positive vortices must have phase steps of $\pi$ half-way between the vortices, which are characteristic of dark solitons, but unlike dark solitons, the phase steps become smaller in the perpendicular direction.

We numerically integrate the GPE~\eref{eqn:nls0} with the potential $V_{\mathrm{trap}} = \frac{1}{4}y^2$. Instead of fully relaxing in imaginary time to any initial state, we print the phase distribution corresponding to a vortex array at $t = 0$ on top of the ground state of $V_{\mathrm{trap}}$, and let the density respond by relaxing for 0.1 (imaginary) time units. We choose $C_{\mathrm{2D}} = 400$, which corresponds to $\xi = 0.4$. The simulation (see~\Fref{fig:samesigns}) confirms that the vortex array initially remains stationary, but once the opposite vortices have appeared, a dark soliton is formed (at $t = 1.4$). The following decay back into vortices swops their original charges.

\section{Conclusions}

We have theoretically shown that the snake instability of (ring) dark solitons can be reversible, if the condensate boundary is chosen appropriately. However, while the decay can be reversible and all the vortices and antivortices disappear, the smooth density distribution of the original RDS might not be reproduced exactly, which makes further revivals less likely to occur. In the limit that the width of the annulus (with respect to the healing length) is taken to zero, we showed that the RDS becomes long-lived.

The snake instability of a RDS has been shown to ultimately result into an example of quantum turbulence~\cite{0953-4075-47-2-021002}. The results are important because they shed light on the coherence of the vortices prior to a fully turbulent state appearing, which prevents further revivals. We have shown that further revivals become less likely the more non-uniformly the vortex-antivortex pairs in the necklace appear after the most recent snake instability. This result raises an interesting notion about the relation between randomness in the positions of the (anti)vortices and occurrence of quantum turbulence.

\ack
We acknowledge the support of the Academy of Finland (grant 133682) and Jenny and Antti Wihuri Foundation (LT).

\section*{References}
\bibliographystyle{unsrt}
\bibliography{../references}
\end{document}